\documentclass[11pt]{article}

\usepackage[nottoc,notlot,notlof]{tocbibind}
\usepackage{amsbsy}
\usepackage{pst-plot}
\usepackage{lscape}
\usepackage{longtable}
\usepackage{rotating}
\usepackage{graphicx}
\usepackage{a4}
\usepackage{hhline}
\usepackage{latexsym}
\usepackage{pifont}
\usepackage{amscd,latexsym,amsmath,euscript,enumerate,verbatim,calc}
\usepackage{mathrsfs}
\usepackage{amsfonts,amssymb,latexsym,amscd,amsmath,euscript,enumerate,verbatim,calc}
\usepackage{hhline}

\pagebreak
\usepackage{tikz}
\usetikzlibrary{arrows,trees}
\usepackage{mathrsfs}
\usepackage{setspace}

\usepackage{amsmath}
\usepackage{amsthm}
\usepackage{amssymb}
\usepackage{graphicx}
\usepackage{natbib}
\usepackage{graphicx,type1cm,eso-pic,color}

\makeatletter
          \AddToShipoutPicture{
            \setlength{\@tempdimb}{.5\paperwidth}
            \setlength{\@tempdimc}{.5\paperheight}
            \setlength{\unitlength}{1pt}
            \put(\strip@pt\@tempdimb,\strip@pt\@tempdimc){
        \makebox(0,0){\rotatebox{55}{\textcolor[gray]{0.85}
        {\fontsize{3cm}{3cm}\selectfont{}}}}
            }
        }
\makeatother

\numberwithin{equation}{section}

\begin{document}
\setcounter{page}{1}
\thispagestyle{empty}
\markboth{}{}

\pagestyle{myheadings}
\markboth{Journal, Indian Statistical Association}{Cause specific rate functions}

\date{}



\vspace{.1in} 

\baselineskip 20truept

\begin{center}
	{\Large {\bf Cause specific rate functions for panel count data with multiple modes of recurrence}} 
\end{center}

\vspace{.1in} 

\begin{center}
	{\large {\bf Sankaran P. G.}} \\
	{\large {\it Cochin University of Science and Technology}} \\
	{\large {\bf Ashlin Mathew P. M.}}\\
	{\large {\it St.Thomas' College (Autonomous), Thrissur}}\\
	{\large {\bf Sreedevi E. P.$^*$}} \\
	{\large {\it SNGS College, Pattambi}} 
\end{center}

\vspace{.1in}
\baselineskip 18truept

\begin{abstract}

	Panel count data arise from longitudinal studies on recurrent events where each subject is  observed only at discrete time points. If recurrent events of several types are possible, we obtain panel count data with multiple modes of recurrence. Such data is commonly encountered in  medical studies, reliability experiments as well as in sociological studies. In this article, we present cause specific rate functions for the analysis of panel count data with multiple modes of recurrence and develop nonparametric estimation procedures for the same. We derive empirical estimators for the cause specific rate functions and also propose a smoothed version of the same estimators using kernel estimation method. Asymptotic properties of the proposed estimators are studied. A simulation study is conducted to assess the performance of the proposed estimators in finite samples. The practical utility of the proposed method is demonstrated using a real life data arising from skin cancer chemo prevention trial given in \cite{sun2013statistical}.
	
\end{abstract}

\vspace{.1in} 


\newpage
\section{Introduction}
\par In survival or reliability studies the events which occur repeatedly are referred to as recurrent events. Examples of recurrent events include occurrences of the hospitalisation of intravenous drug users \citep{wang2001analyzing}, repeated occurrence of some  tumours, and the warranty claims for automobiles \citep{kalbfleisch1991methods}. Recurrent events have been further classified based on the monitoring patterns. If the study subjects are monitored continuously and the information on the occurrence times of all events are available, the data is termed as recurrent event data \citep{cook2007statistical}. When the study subjects are examined or observed only at discrete time points, may be because the continuous observation is too expensive or unable to obtain, such data is referred to as panel count data (\citet{kalbfleisch1985analysis}, \citet{sun2009analyzing}, \citet{zhao2011nonparametric}). Panel count data is also termed as interval count data or interval censored recurrent event data (\cite{lawless1998analysis} and \cite{thall1988analysis}). If each subject is observed only once, a special case of panel count data arises which is commonly known as current status data.

\par The analysis of panel count data focuses mainly on the rate function and mean function of the underlying recurrent event process. \cite{thall1988analysis}  and \cite{lawless1998analysis}  considered the analysis of panel count data using rate functions. An estimator for the mean function based on isotonic regression theory was developed by \cite{sun1995estimation}. \cite{wellner2000two} discussed likelihood  based nonparametric estimation methods for the mean function and proposed a nonparametric maximum likelihood estimator (NPMLE) and a nonparametric maximum pseudo likelihood estimator (NPMPLE) for the same. They also showed that NPMPLE is exactly the one studied in \cite{sun1995estimation}. Some of important research works in this area include \cite{wellner2007two}, Huang et al. (2006), Hu et al. (2009), \cite{xu2018joint} and \cite{chiou2019semiparametric} and references there in.


\par When an  individual (subject) in the study is exposed to the risk of multiple types of events at each point of observation, we obtain panel count data with competing risks or multiple modes of recurrence. Such data naturally arise from survival and reliability studies where the interest is focused on the recurrence of competing events which can be observed only at discrete time points. Recurrent event data with competing risks is studied by many authors in literature \citep{cook2007statistical}. However,  sparse literature only exists on the analysis of panel count data with multiple modes of recurrence. Recently, \cite{sreedevi2020nonparametric}  derived an expression for the cause specific mean functions and developed a nonparametric test for comparing the effect of different causes on recurrence time based on the developed estimators. To the best of our knowledge, no literature exists on rate functions for panel count data with multiple modes of recurrence. There are several advantages in using rate functions for the analysis of panel count data compared to mean functions. Mainly, fewer assumptions are  only required for models based on rate functions. In addition, rate functions are not constrained with the non decreasing property of mean functions and hence it is easy to understand the changing recurrence patterns with rate functions. Rate functions  can be used for the graphical representation of the underlying process of a panel count data as the hazard rate functions for failure time processes. We can also note that  an estimator of rate function could be used to derive the estimator of  corresponding mean function. However, due to the difficulty in deriving asymptotic properties of the rate functions, it is less explored in literature. This motivated us to study the features of rate functions of panel count data when individuals (subjects) are exposed to several types of events which may recur and to develop estimation procedures for the same. 

\par The paper is organized as follows. We present cause specific rate functions for panel count data with multiple modes (causes) of recurrence and discuss their estimation in Section 2. In Section 3,  we derive empirical estimators for cause specific rate functions.  Further, a smoothed version of the estimators based on kernel estimation technique is developed. We use Gaussian kernel to smooth the estimates of cause specific rate functions. Large sample properties of the kernel estimators are  discussed. An extensive simulation study is carried out in Section 4 to assess the finite sample performance of the proposed estimators. The estimators developed in this study are applied to a real data on skin cancer chemo prevention trials in Section 5. Finally, Section 6 summarizes the major conclusions of the study with a discussion on future works. 
\vspace{.1in} 
\section{Cause specific rate functions}
\par We introduce cause specific rate functions for the analysis of panel count data with multiple modes of recurrence in this section.   We also discuss various estimators of cause specific rate functions.

Consider a  study on $n$ individuals from a homogeneous population which are exposed to the recurrent events due to $\{1,2,...,J\}$  possible causes. Assume that the event process is observed only at a sequence of random monitoring times. Consequently, the counts of the event recurrences due to each cause  in between the observation times are only available; the exact recurrence times remain unknown. Accordingly, we observe the  cumulative number of recurrences upto every observation time due to each cause. Define a counting process $ N_j=\{N_j(t); t\ge 0\}$ where $N_j(t)$ denote the number of recurrences of the event due to cause $j$ upto time $t$.
Define $\mu_j(t)=E(N_j(t))$ as the mean function of the recurrent event process due to cause $j$ which are termed as cause specific mean functions. Define $r_j(t)dt=d\mu_j(t)=EdN_j(t)$ as the rate function of the recurrent event process due to cause $j$, for  $j=1,2,...,J$. We now refer $r_j(t)$ as the cause specific rate function, which is not yet studied in literature till date. By studying cause specific rate functions, one can easily understand the difference in recurrence patterns due to various causes (modes) of recurrence. 
The goal of this study is to develop  nonparametric estimation procedures for cause specific rate functions $r_j(t)$, $j=1,2,...,J$.  

\par Note that the number of observation times as well as observation time points may be different for each individual.  Let $M_{i}$ be an integer valued random variable denoting the number of observation times for $i=1,2,..,n$. Also let $T_{i,p}$ denote the $p^{th}$ observation time for $i^{th}$ individual for $p=1,2,..M_i$ and $i=1,2,..,n$. Assume that the number of recurrences due to different causes are independent of number of observation times as well as observation time points.  Let $N_{i,p}^{j}$ denote the number of recurrences of the event observed  for $i^{th}$ individual due to cause $j$ , for $p=1,2,...,M_{i}$, $i=1,2,...,n$ and $j=1,2,...,J$. Now  we observe $n$ independent and identically distributed copies of  $\{M_{i},T_{i,p},N_{i,p}^{1},...,N_{i,p}^{J}\}$, $p=1,2,...,M_{i}$.
The observed data will be of the form $\{m_{i},t_{i,p},n_{i,p}^{1},...,n_{i,p}^{J}\}$, $p=1,2,...,m_{i}$ and $i=1,2,...,n$. 
Let $b_{1}<b_{2}<...<b_{l}$ are the distinct observed time points in the set $\{T_{i,p}, ~p=1,2,...,M_{i},~ i =1,2,...,n$\}. Since $r_j(t)dt=d\mu_j(t)$, possible estimators of $r_j(t)$'s are
\begin{equation}
\widehat r_{1j}(b_q)=\hat\mu_j(b_q)-\hat\mu_j(b_{q-})~~~~q=1,2,...,l~~\text{and}~~j=1,2,...,J
\end{equation}
where $\widehat r_{1j}(t) =0$ for all other $t \neq b_q$. Now (2.1) can be modified as 
\begin{equation}
\widehat r_{2j}(t)=\frac{\Delta\hat\mu_j(b_{q})}{b_{q}-b_{q-1}}~~~~b_{q-1}<t < b_{q},~~q=1,2,...,l~~\text{and}~~j=1,2,...,J.
\end{equation}
In Eqns (2.1) and (2.2), $\hat\mu_j(t)$, $j=1,2,...,J$ are the estimators of cause specific mean functions developed in Sreedevi and Sankaran (2020). One disadvantage in using these estimators is that, the estimators of cause specific mean functions are required in advance.\\
\par Under the Poisson assumption of independence of increments for underlying recurrent event processes, maximum likelihood estimators for cause specific rate functions can be derived. Maximum likelihood estimators for $r_j(t)$,~$j=1,2,...,J$ can be obtained by extending the results in non competing risks set up discussed by  Wellner et al. (2004, pp:149). Cause specific rate functions can also be estimated by assuming them as piece wise constant functions in predefined time intervals and using likelihood based estimation procedures (Sun and Zhao (2013), pp:63). Both of these likelihood based estimators do not have explicit forms which make the computation tedious. Motivated by this, we  introduce empirical estimators for cause specific rate functions which can be directly estimated from the given data. The kernel based smoothing for the empirical estimators are also discussed. The advantage of using empirical estimators is that they can be expressed in explicit forms and the computation is very easy. 

\section{Empirical and kernel estimators for cause specific rate functions}
In this section, we first derive empirical estimators for cause specific rate functions. We can see that the estimators of cause specific rate functions have jumps at observed distinct time points. This motivates us to propose smooth estimators for cause specific rate functions using kernel estimation technique. We also discuss the asymptotic properties of the proposed kernel estimators.
\subsection{Empirical estimators for cause specific rate functions}
\par Consider the data structure of panel count data with multiple modes of recurrence which is described in Section 2. The observed data is given in the form $\{m_{i},t_{i,p},n_{i,p}^{1},...,n_{i,p}^{J}\}$,  $p=1,2,...,m_{i}$ and $i=1,2,...,n$. We now propose empirical estimators for the cause specific rate functions $r_j(t)$, which is the average of empirical rate functions due to cause $j$ over all individuals. Define 
\begin{equation}
	\widehat{r_j(t)}=\frac{\sum_{i=1}^{n}\left[\sum_{p=1}^{m_i}\frac{
			(n_{i,p}^{j}  - n_{i,p-1}^{j})   I(t_{i,p}  < t \leq t_{i,p-1} )}{(t_{i,p} - t_{i,p-1}  )}\right]}{\sum_{i=1}^{n}(t\le t_{i,p} )}~~~~j=1,2,...,J.
	\label{eq1}
\end{equation}
In this definition, the numerator gives the average  number of recurrences for subject $i$  due to cause $j$ and denominator is the number of individuals at risk at time $t$. Hence the estimators $\widehat {r_j(t)}$'s are the average of rate functions due to cause $j$ over all individuals.  We can  estimate the cause specific rate functions directly from Eqn (3.1). When $J=1$, $\widehat{r_j(t)}$ reduce to the estimator of rate function studied by Sun and Zhao (2013) in non competing risks set up. Thall (1988) and Thall and Lachin (1988) studied the estimators of rate functions for panel count data in non competing risks set up. The plots of the empirical cause specific rate functions  provide a basis for determining a reasonable form for the underlying rate as a function of time. The data of time of recurrence and number of recurrences can be completely reconstructed from the graph of $\widehat {r_j(t)}$ for cause $j$, as it contains all of the available information. A $100(1-\alpha)\%$ confidence interval for $\widehat{r_j(t)}$ can be computed using standard methods (Thall and Lachin, 1988).


\subsection{Kernel estimators of cause specific rate functions}
	
\par In practice, we can see that the estimators of cause specific rate functions presented in Eqn (\ref{eq1}) are changing only at the observed time points. This motivated us to propose a smoothed version of the estimators for cause specific rate functions using kernel estimation techniques. We also study the asymptotic properties of kernel estimators. The kernel estimators we propose are weighted averages of the estimators of cause specific rate functions given in Eqn (\ref{eq1}). 
Let $K(t)$ be a non-negative kernel function symmetric about $t=0$ with $\int_{-\infty}^{\infty}K(t)dt=1$. Also let $h_n>0$ be the bandwidth parameter. Let $b_1<b_2<...<b_l$ denote the distinct ordered time points where $r_j(t)$ is estimated and define $\widehat{r_{qj}}=\widehat{r_j(b_q)}$, for $q=1,2,...,l$, $j=1,2,...,J$. Define 
\begin{equation}
	w^*_q(t,h_n)=h_n^{-1}K\left(\frac{t-b_q}{h_n}\right)
\end{equation}
	and 
\begin{equation}
	w_q(t)=\frac{w^*_q(t,h_n)}{\sum_{u=1}^{l}w^*_u(t,h_n)} ~~~~  q=1, 2,...,l.
\end{equation}
Now, the kernel estimators of $r_j(t)$'s are given as
\begin{equation}
\widehat{r^{*}_{j}(t)}= \sum_{q=1}^{l} w_q(t)  \widehat{r_{qj}}~~~~j=1,2,...,J.
\end{equation}
We can see that the smoothed estimators $\widehat{r^{*}_{j}(t)}$ of the cause specific rate functions are weighted average of $\widehat{r_{j}(t)}$'s. Many choices for kernel functions are there in literature. In this study we choose the Gaussian kernel given by 
\begin{equation}
	K(t)=(2\pi)^{-1/2}\text{exp}(-t^{2}/2).
\end{equation}
While using Gaussian kernel function, all components of $\widehat{r_{j}(t)}$'s contribute to their resulting estimators at each time point. The amount of contribution depends on the closeness  of each time point to the given $t$ and the closer, the larger the contribution (Sun and Zhao, 2013). The bandwidth for which the MSE is minimum is selected to employ smoothing. 
\par As we stated earlier the asymptotic properties of the estimators of rate functions of panel count data are not developed yet. However, the asymptotic properties of the estimators $\widehat{r^{*}_{j}(t)}$'s are studied and we derive the following results.
Without loss of generality, we can assume that the kernel function $K(x)$ satisfies the following conditions.\\
C1 : $K(x)$ is bounded ie sup\{$K(x), x \in R \}<\infty $\\
C2 : $|xK(x)| \to 0$ as $|x| \to \infty$  \\  
C3 : $K(x)$ is symmetric about 0, ie $K(-x)=K(x)$, $x \in R$\\
Also suppose that, as $n \to \infty$ the bandwidth parameter $h_n$ satisfies the conditions  
(i) $h_n \to 0$ (ii) $nh_n \to \infty$ and (iii) $nh^{2}_n \to \infty$. \\
\textbf{Result 1}: Under the assumptions C1, C2 and C3, the estimators  $\widehat{r^{*}_{j}(t)}$'s are asymptotically unbiased estimator of $r_j(t)$'s for  every fixed $t$ at which $r_j(t)$'s are defined and continuous
\begin{equation}
    E(\widehat{r^{*}_{j}(t)})={r_{j}(t)}~~\text{as}~~~ n\to \infty~~ j=1,2,...,J.
\end{equation}
\textbf{Result2}: Under the assumptions C1, C2 and C3, the estimators  $\widehat{r^{*}_{j}(t)}$'s are consistent estimators in quadratic mean  of ${r_{j}(t)}$'s  for every fixed $t$ at which $r_j(t)$'s are defined and continuous 
\begin{equation}
	E(\widehat{r^{*}_{j}(t)}-{r_{j}(t)})^{2} \to 0~~\text{as}~~~ n\to \infty~~j=1,2,...,J.
\end{equation}
\textbf{Result3}: Under the assumptions C1, C2 and C3, for fixed $t$, the estimators  $\widehat{r^{*}_{j}(t)}$'s  are asymptotically normal with mean $\lambda_j(t)=E(\widehat{r^{*}_{j}(t)})$ and standard deviation $\sigma_j(t)=\text{s.d}(\widehat{r^{*}_{j}(t)})$ for $j=1,2,...,J$.

The above results can be easily verified from Theorems 5, 6 and 7 of Section 15.5.1 in \cite{roussas2003introduction}.
		
\section{Simulation studies}

Simulation studies are conducted to  assess the performance of the proposed estimators of the cause specific rate functions in finite samples. We consider the situation with two competing risks. 
The real life situations in reliability and survival studies are taken as a model to generate panel count data of the form 
$\{m_{i},t_{i,p},n_{i,p}^{1},n_{i,p}^{2}\}$ for $p=1,2,...,m_{i}$ and $i=1,2,...,n$. The number of observation times $m_{i}$ for each individual is generated from a discrete uniform distribution $U(1,10)$ for $i=1,2,...,n$. Thus the maximum number of observations for each individual is restricted upto 10. Then we generated gap times between each observation from uniform distribution $U(0,5)$. The discrete observation time points $t_{i,p}$ for $p=1,2,...,m_i$ and $i=1,2,...,n$ are generated using the above mentioned time gaps. A bivariate Poisson distribution with parameters $(\theta_1, \theta_2, \theta_3)$ is employed to generate recurrent processes $n_{i,p}^{1}$ and $n_{i,p}^{2}$. The joint mass function of the bivariate Poisson distribution with parameters $(\theta_1, \theta_2, \theta_3)$ is given by 
\begin{equation}
f(x,y)=\exp\{-(\theta_1+\theta_2+\theta_3)\} \frac{{\theta_1} ^x}{x!}\frac{{\theta_2}^y}{y!}\sum_{k=0}^{min(x,y)}{x \choose k}{y \choose k}k! \left(\frac{\theta_3}{\theta_1 \theta_2}\right)^k.
\end{equation}
\par 
Marginally each random variable follows Poisson distribution with $E(X)=\theta_{1}+\theta_{3}$, $E(Y)=\theta_{2}+\theta_{3}$  and cov$(X,Y)=\theta_{3}$ gives a measure of dependence between random  variables $X$ and $Y$. In a non-competing risks set up, \cite{balakrishnan2011class} used a similar procedure to generate panel count data.

\begin{table}[]
	\caption{ Absolute bias and MSE of $\widehat{r_{j}(t)}$ for $j=1,2$ at different time points}
	\centering
	\begin{tabular}{|c|c|c|c|c|}
		\hline
		& \multicolumn{2}{c|}{$\widehat{r_{1}(t)}$} & \multicolumn{2}{c|}{$\widehat{r_{2}(t)}$}\\ \hline
		Times & Absolute Bias    & MSE       & Absolute bias    & MSE       \\ \hline
		
		\multicolumn{5}{|c|}{n= 100}                                        \\ \hline
		1     & 0.0776           & 0.0063    & 0.0561            & 0.0316     \\ 
		2     & 0.0059           & 0.0006    &  0.0384           & 0.0148     \\ 
		3     & 0.0325           & 0.0015    &  0.0283            & 0.0807    \\ 
		7     & 0.0739           & 0.0062    & 0.0140           & 0.0200      \\	
		8     & 0.0845           & 0.0089    &  0.0147           & 0.0222    \\ 
		9     & 0.0977           & 0.0117    & 0.0963           & 0.0276    \\ 
		10    & 0.1501           & 0.0253    & 0.0620           & 0.0153    \\ 
		13    & 0.0485           & 0.0248    & 0.0382           & 0.0149     \\ 
		15    & 0.0889            & 0.0827    & 0.0961           & 0.0380    \\ 
		16    & 0.0912           & 0.0871    & 0.0506           & 0.0260    \\  \hline
		\multicolumn{5}{|c|}{n = 200}                                       \\ \hline
		1     & 0.0682           & 0.0053    & 0.0540            & 0.0295    \\ 
		2     & 0.0044           & 0.0004    &  0.0364           & 0.0136    \\ 
		3     & 0.0212           & 0.0008    &  0.0282          & 0.0805  \\ 	
		7     & 0.0515           & 0.0034    & 0.0628           & 0.0042    \\ 
		8     & 0.0763           & 0.0067    & 0.0106           & 0.0120     \\
		9     & 0.0545           & 0.0037    & 0.0859           & 0.0077    \\ 
		10    & 0.0949           & 0.0101    & 0.0529           & 0.0042    \\ 
		13    & 0.0431           & 0.0127      & 0.0219           & 0.0103    \\ 
		
		15    & 0.0812           & 0.0083    & 0.0941           & 0.0095    \\ 
		16    & 0.0757           & 0.0073    & 0.0129           & 0.0172     \\ \hline
		\multicolumn{5}{|c|}{n = 500}                                       \\ \hline
		
		1     & 0.0512        & 0.0031     & 0.0523            & 0.0274    \\ 
		2     & 0.0029           & 0.0003    & 0.0359           & 0.0130   \\ 
		3     & 0.0195           & 0.0006    &   0.0239            & 0.0584   \\ 
	 	7     & 0.0251           & 0.0013    & 0.0288           & 0.0012    \\ 
		8     & 0.0038            & 0.0006    & 0.0099           & 0.0093    \\ 
	    9     & 0.0097           & 0.0006    & 0.0343           & 0.0014    \\ 
		10    & 0.0287           & 0.0016    & 0.0404           & 0.0018    \\ 
    	13    & 0.0372           & 0.0023   &  0.0181           & 0.0016    \\
		15    & 0.0517           & 0.0037    & 0.0643           & 0.0091     \\ 
		16    & 0.0587           & 0.0044    & 0.0113           & 0.0130     \\  \hline
	\end{tabular}
\end{table}

\begin{table}
	
	\caption{ Absolute bias and MSE of $\widehat{r^{*}_{j}(t)}$ for $j=1,2$ at different time points}
	\centering
	\begin{tabular}{|c|c|c|c|c|} 
		\hline
		& \multicolumn{2}{c|}{$ \widehat{r^{*}_{1}(t)}$} & \multicolumn{2}{c|}{$\widehat{r^{*}_{2}(t)}  $}\\ \hline
		Times & Absolute Bias    & MSE       & Absolute bias    & MSE       \\ \hline
		\multicolumn{5}{|c|}{n = 100}                                        \\ \hline
		1     & 0.0624          & 0.0047          & 0.0560           & 0.0315  \\     
		
		2       & 0.0280         & 0.0019         & 0.0439        & 0.0194  \\ 
		
		3     & 0.0340          & 0.0016               & 0.0381           & 0.0146  \\ 
		
		6     & 0.0689          & 0.0054              & 0.0231           & 0.0542  \\ 
		
     	7     & 0.0880          & 0.0085               & 0.0219         & 0.0486  \\ 
		
		8     & 0.1517         & 0.0242               & 0.0257          & 0.0665  \\ 
		
		9     & 0.0967          & 0.0103             & 0.0172          & 0.0300  \\ 
		
		10    & 0.1683         & 0.0299            & 0.0186           & 0.0353 \\

		13    & 0.2199          & 0.0526               & 0.0412           & 0.0030   \\ 
		
		16    & 0.2491          & 0.0680         & 0.1671          & 0.0281  \\  
		\hline
		& \multicolumn{1}{l}{} & \multicolumn{1}{l}{n=200}  & \multicolumn{1}{l}{} &               \\ 
		\hline
		1     & 0.0585         & 0.0038               & 0.0534     & 0.0287   \\ 
		
		2    & 0.0248         & 0.0013                  & 0.0418          & 0.0175 \\ 
		
		3     & 0.0078        & 0.0007           & 0.0375           & 0.0141  \\ 
		
		6     & 0.0625         & 0.0052            & 0.0175      & 0.0317   \\ 
		
		7     & 0.0786        & 0.0076             & 0.0154           & 0.0246   \\ 
		
		8     & 0.0827         & 0.0083             & 0.0119         & 0.0149  \\ 
		
		9     & 0.0666         & 0.0061              & 0.0029         & 0.0101   \\ 
		
		10    & 0.1066          & 0.0140               & 0.0112           & 0.0128   \\ 
		
		13    & 0.0987         & 0.0110            & 0.0044           & 0.0023   \\  
		
		16    & 0.1542          & 0.0260          & 0.1090        & 0.0119 \\         
		\hline
		& \multicolumn{1}{l}{} & \multicolumn{1}{l}{n= 500} & \multicolumn{1}{l}{} &               \\ 
		\hline
		1     & 0.0462          & 0.0023              & 0.0521          & 0.0272  \\ 
		
		2       &  0.0025          & 0.0003   & 0.0381          & 0.0146  \\
		
		3     & 0.0023          & 0.0001          & 0.0308          & 0.0095  \\ 
		
		6     & 0.0044          & 0.0002               & 0.0073        & 0.0054  \\ 
		
		7     & 0.0012          & 0.0002              & 0.0030          & 0.0010 \\ 
		
		8     & 0.0040          & 0.0002               & 0.0015          & 0.0003  \\ 
		
		9     & 0.0161          & 0.0004              & 0.0291           & 0.0009  \\ 
		
		10    & 0.0077         & 0.0002              & 0.0062          & 0.0039 \\ 
		
		13    & 0.0036        & 0.0003              &  0.0036         & 0.0016  \\ 
		
		16    & 0.0180         & 0.0005            &  0.0018          & 0.0094  \\   
		\hline
	\end{tabular}
\end{table}

\par The absolute bias and mean square error (MSE) of the estimates of cause specific rate functions at 10 randomly chosen observation points are calculated. For this purpose 1000 random samples of sizes, $n= 100, 200$ and $500$ are generated. The absolute bias and MSE are  calculated at randomly chosen time points $1,2,3,7,8,9,10,13,15,16 $. Various parameter combinations of $(\theta_1,\theta_2,\theta_3)$ are used to generate the number of recurrences. As the results are similar, here we present the same for $(\theta_1,\theta_2,\theta_3)=(0.2,0.3,0.5)$. The absolute bias and MSE of $ \widehat{r_{j}(t)}$ for $j=1,2$  are given in Table 1. 

Different choices of bandwidths are used to employ smoothing using kernel estimation. The optimum bandwidth
is selected as $h_n=n^{\frac{1}{10}}$, which minimizes MSE. The absolute bias and MSE of $ \widehat{r^{*}_{j}(t)}$ for $j=1,2$ calculated using bandwidth value $h_n=n^{\frac{1}{10}}$ are presented in Table 2.  From Table 1 we can infer that both absolute bias and MSE  decrease with the increase in the sample size. Same results are observed from Table 2 also. Further, we note that the absolute bias and MSE values of the estimators does not vary monotonically with time points. It can be noted that there is not much difference between the absolute bias and MSE of the proposed estimators and the corresponding kernel estimators, which shows both estimators perform equally good in finite samples. Hence both estimators can be used in a practical situation. However the asymptotic properties of kernel estimators were established in Section 3.

\section{Data analysis}
The  proposed estimators are applied to a real data on  skin cancer chemo prevention trial given in \cite{sun2013statistical} for illustration.
The main objective of this study was to evaluate the effectiveness of the DFMO (DIfluromethylornithire) drug in reducing new skin cancers in a population with a history of non-melanoma skin cancers, basal cell carcinoma and squamous cell carcinoma. The data consists of 290 patients with a history of non-melanoma skin cancers. The real observation and follow up times differ for each patient. The data has the counts of two types of recurring events basal cell carcinoma and squamous cell carcinoma which we treat here as two competing risks \citep{sreedevi2020nonparametric}.

In the data set, the number of observations on an individual varies from 1 to 17 and the time of observation varies from 12 to 1766 days. The cause specific rate functions due to basal  cell carcinoma and squamous cell carcinoma are estimated using Eqn (\ref{eq1}). The plots of the estimated cause specific rate functions are given in Figure 1. Further, kernel estimators with different bandwidths are used to smooth the estimator. The plots of the kernel estimators with bandwidth parameter value $h_n=1.76 \approx n^{\frac{1}{10}} $ is given in Figure 2. For different bandwidths, we can see that as $h_n$ increases the smoothness of the curve also increases.

\begin{figure}[!htb]
	\centering
	\includegraphics[width=0.9\linewidth]{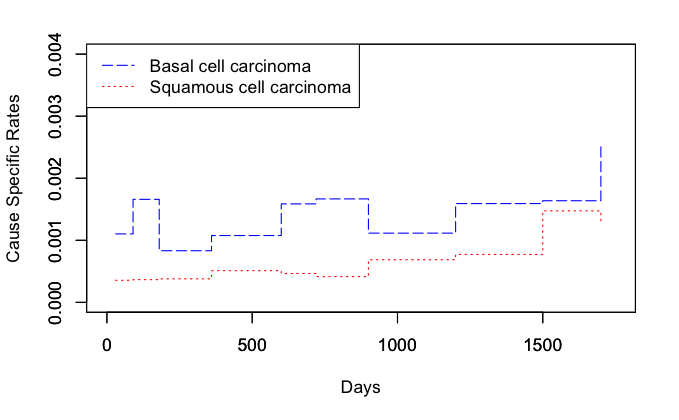}
	\caption{Empirical estimates of cause specific rate functions due to basal cell carcinoma and squamous cell carcinoma}
	\label{fig:skintumor}
\end{figure}

\begin{figure}[!htb]
	\centering
	\includegraphics[width=0.9\linewidth]{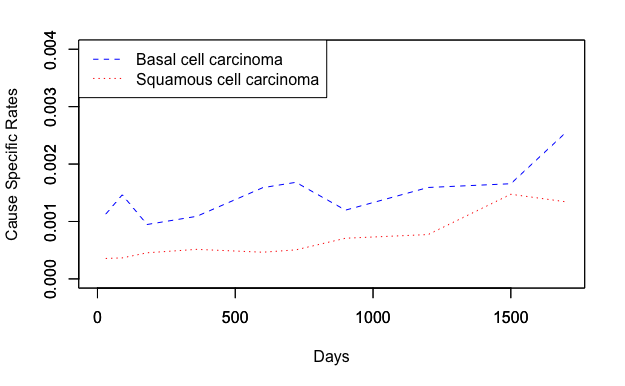}
	\caption{Kernel estimates of cause specific rate functions  due to basal cell carcinoma and squamous cell carcinoma for $h_n=1.76$ }
	\label{fig:skintumor2}
\end{figure}

\par From Figure 1, it can be noted that the recurrence rate of basal cell carcinoma is greater than the recurrence rate of squamous cell carcinoma at all time points. Since the rate functions are not monotonic, the change points of recurrence patterns can be easily identified from the graph. It is clear from Figure 1 that, the recurrence pattern of cause specific rate function due to basal cell carcinoma and squamous cell carcinoma are entirely different. For example, at time point near to 950 days, the cause specific rate function due to basal cell carcinoma decreases, while the same due to squamous cell carcinoma increases. 
\begin{figure}[!htb]
	\centering
	\includegraphics[width=0.9\linewidth]{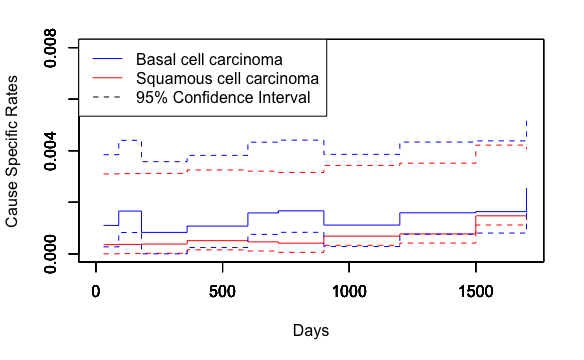}
	\caption{Empirical estimates of cause specific rate functions due to basal cell carcinoma and squamous cell carcinoma with 95\% confidence interval  }
	\label{fig:skintumor}
\end{figure}

\begin{figure}[!htb]
	\centering
	\includegraphics[width=0.9\linewidth]{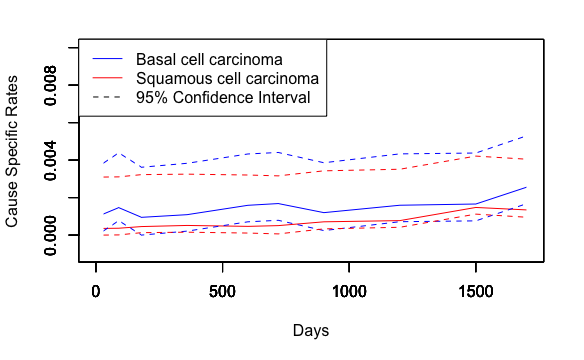}
	\caption{Kernel estimates of cause specific rate functions  due to basal cell carcinoma and squamous cell carcinoma for $h_n=1.76$ with 95\% confidence interval}
	\label{fig:skintumor2}
\end{figure}
The standard procedures are used to estimate the 95\% confidence intervals for the cause specific rate functions. The estimates of cause specific rate functions due to basal cell carcinoma and squamous cell carcinoma  with 95\% confidence intervals are plotted in Figure 3 and Figure 4. In both Figures 3 and 4 , solid line represents the actual estimates and the dashed line represents the 95\% confidence intervals.

\section{Conclusion}

\par In this paper, we studied cause specific rate functions and their nonparamertic estimation for the analysis of panel count data with multiple modes of recurrence. We introduced empirical estimators for cause specific rate functions which have simple closed forms. The interpretation of the proposed estimators is straightforward and calculation is simple. Further kernel estimation procedure using Gaussian kernel is employed to smooth the estimators of cause specific rate functions. An extensive simulation study is carried out by generating data from bivariate Poisson process to assess the finite sample performance of the proposed estimators. The proposed methods are illustrated using a real data on skin cancer chemo prevention trials given in \cite{sun2013statistical}.
\par We can develop likelihood based estimators for cause specific rate functions as mentioned in Section 2. However, those estimators do not have a closed from expression and numerical methods are needed for the estimation procedure. Study in this direction will be reported in a separate research work. When the study subjects are exposed to the risk of recurrence due to several possible causes (modes), it is important to know whether all causes effect recurrence time identically. We can compare cause specific rate functions due to different modes of recurrence to  address  this problem when panel count data is only available. The work will be reported elsewhere. An area of recent interest in panel count data analysis is to use M-spline and B-spline functions. The analysis of panel count data with multiple modes of recurrence using such nonparametric methods are yet to be studied. Regression methods for panel count competing risks data are also under investigation.

\begin{center}
{\bf Acknowledgments}\\
\end{center}
\par We thank the editor and unknown reviewers for their constructive comments which help to improve this research work. The first author would like to thank Science Engineering and Research Board, DST, Government of India and the third author acknowledge the gratitude to Kerala State Council for Science Technology and Environment for the financial support provided to carry out this research work.

\vspace{.2in} 
\newpage
\noindent 
{\bf Sankaran P. G.} \\
Department of Statistics\\
Cochin University of Science and Technology, Kochi-22\\
E-mail: sankaran.p.g@gmail.com\\

\vspace{.1in}
  
\noindent 
{\bf Ashlin Mathew P. M.}\\  
Department of Statistics\\
St. Thomas' College(Autonomous), Thrissur-1\\
E-mail: ashlinmathewpm@gmail.com \\

\vspace{.1in}

\noindent 
{\bf Sreedevi E. P.}$^*$\\  
Department of Statistics\\
SNGS College, Pattambi, Palakkad\\
E-mail: sreedeviep@gmail.com \\
\end{document}